\documentclass[10pt,twocolumn,a4paper]{article}
\usepackage[T1]{fontenc}
\usepackage[cp1250]{inputenc}
\usepackage{graphicx}
\usepackage{epstopdf}
\usepackage{lmodern}

\usepackage{varioref}% nice with \autoref{label}, use \vpageref[]{label}
\usepackage{amsmath,amssymb,amsfonts}

\usepackage[pdftex,breaklinks=true]{hyperref}
\hypersetup{
pdfauthor={Piotr Bania},
pdfstartview=FitH,
pdfsubject={Security Mitigations for Return-Oriented Programming Attacks},
pdftitle={Security Mitigations for Return-Oriented Programming Attacks}
}

\usepackage{breakurl}
\usepackage{url}
\usepackage{listings}
\usepackage{color}

\title{Security Mitigations for Return-Oriented Programming Attacks}
\author{Piotr Bania\\Kryptos Logic Research\\
\texttt{\href{http://www.kryptoslogic.com}{www.kryptoslogic.com}}}
\date{2010}

\begin{document}
\maketitle

\begin{abstract}

With the discovery of new exploit techniques, new protection mechanisms are needed as well. Mitigations like DEP (Data Execution Prevention) or ASLR (Address Space Layout Randomization) created a significantly  more difficult environment for vulnerability exploitation. Attackers, however, have recently developed new exploitation methods which are capable of bypassing the operating system's security protection mechanisms.

In this paper we present a short summary of novel and known mitigation techniques against return-oriented programming (ROP) attacks. The techniques described in this article are related mostly to x86-32\footnote{Some of the techniques can be also applied on other architectures, albeit some of them are only available for x86-32 family (e.g., the ones based on creating new segment descriptors).} processors and Microsoft Windows operating systems.

\end{abstract}

\section{Introduction}

In order to increase the security level of the operating system, Microsoft has implemented several mitigation mechanisms, such as DEP and ASLR. Data Execution Prevention (DEP) is a security feature that prohibits the application from executing code from non-executable memory area. To exploit a vulnerability, an attacker must find a executable memory region and be able to fill it with necessary data (e.g., shellcode instructions). Generally, achieving this goal using old exploitation techniques is made significantly more difficult with the addition of the DEP mechanism. As a result, attackers improved upon the classic ``return-into-libc'' technique and started using return-oriented programming (ROP)~\cite{RopPresent,RopPaper} to bypass Data Execution Prevention.

Techniques like ROP are still based on the attacker understanding memory layout characteristics, leading Microsoft to implement Address Space Layout Randomization (ASLR) as a countermeasure. ASLR renders the layout of an application's address space less predictable because it relocates the base addresses of executable modules and other memory mappings. In order to bypass DEP protection mechanism ROP technique was introduced. In this article we present novel and known mechanisms which are created specifically to prevent attackers from exploiting vulnerabilities based on the ROP method. Presented mitigations will be divided in two general categories:

\begin{itemize}
    \item Compiler-level mitigations --- mitigations that can be only applied by the compiler or linker.
    \item Binary-level mitigations --- mitigations that can be applied without knowing the source code of the protected code fragment.
\end{itemize}

\section{Return-oriented Programming}

Return-oriented programming is a known exploitation technique which allows the attacker to use stack memory to indirectly execute previously picked instructions (so called gadgets). Typically each gadget ends with the x86 subroutine return instruction\footnote{However other instructions may be used as well like {\tt{jmp reg}}, {\tt{call reg}} etc.} ({\tt{RET}}), which further transfers the execution to the next gadget or the payload itself. For more information regarding the return-oriented programming technique please refer to~\cite{ROPDino,RopPresent,RopPaper}.

\section{Compiler-level mitigations}\label{sec:compiler_level}

In this section we present ROP protection mechanisms which can be applied at the compiler-level. However this doesn't mean they are not implementable at the binary-level - they are simply substantially easier to implement at the compiler-level. We will also try to underline advantages and disadvantages of described mechanisms.

The biggest disadvantage of compiler-level mitigations is the fact that they require code recompilation in order to be effective. It is often hard to quickly implement such kind of changes in the real world.

\subsection{Call-Ret relations}

\begin{figure}[tbhp]
\centering
\includegraphics[scale=0.7, angle=90]{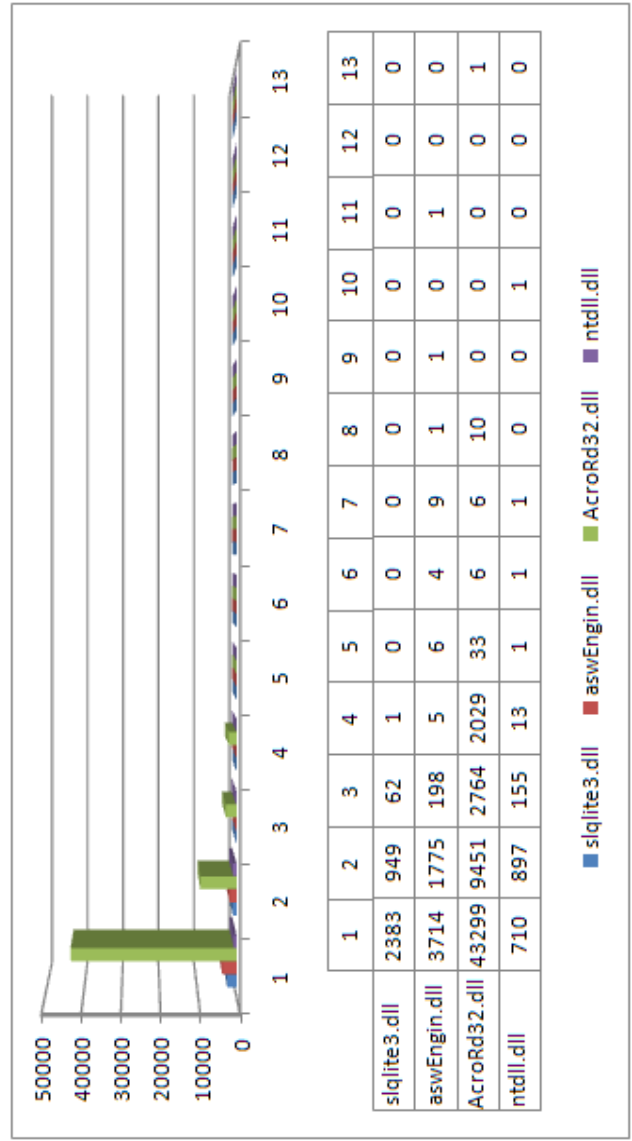}
\caption{{\tt{RET}} opcode offsets in sample modules (offset equal to 1 indicates that this is an original {\tt{RET}} instruction).}
\label{img:wykres}
\end{figure}

As previously stated, most gadgets use return instructions to transfer execution control to another gadget or payload. In order to find useful gadgets, attackers scan the process memory or the binary module for return instruction opcodes and, after such opcode is found, they try to perform backward disassembly in order to decide whether following gadget is useful (correct) or not. Return instruction opcodes can often be found in the middle of different instructions. Results, however, show that most of the time original return instructions {\tt{RET}} are used. Typically they also represent the highest number of return opcodes found in the entire module's executable area (cf. Figure \ref{img:wykres}). For the remainder of this article {\tt{RET}} instructions emitted in the original program's code will be named as ``original return instruction''.\newline

\subsubsection{Testing for CALLs}\label{testing_calls}

In typical applications, every procedure (function) is executed by using call-procedure instruction. Every {\tt{CALL}} instruction saves procedure linking information on the stack and branches to the procedure specified by the destination operand. Our ROP mitigation technique relies on a fact that each return address popped from the stack by the {\tt{RET}} instruction is preceded by {\tt{CALL}} instruction. When a ROP attack occurs the return address points to another gadget (or finally a payload). It is unlikely that an attacker will be able to pick the return addresses preceded by {\tt{CALL}} instruction operands (see \autoref{table:call_results} for details). Testing for {\tt{CALL}} instructions located before the return address popped from the stack should be a reliable method against ROP attacks.

\begin{table}[tbhp]
\centering
\begin{tabular}{| l | c  | c  |}
\hline
Module Name & $N_1$ [\#] & $N_2$ [\#]\\ \hline
{\tt{ntdll.dll}} & 6528 & 138 (2.11\%) \\
{\tt{ieframe.dll}} & 45232 & 2109 (4.66\%)\\
{\tt{bib.dll}} & 5966 & 317 (5.31\%)\\
{\tt{aswEngin.dll}} & 50895 & 1547 (3.03\%)\\
\hline
\end{tabular}
\caption{Number of gadgets preceded by relative, memory indirect, register indirect procedure-call instruction ("minimal/not extended" addressing mode assumed).}
\label{table:call_results}
\end{table}
Where:
\begin{itemize}
    \item $N_1$ represents total number of gadgets
    \item $N_2$ represents number of gadgets preceded by the procedure-call instruction
    \item gadget represents a valid single instruction or sequence of instructions without any special filtering applied regarding the gadget usefulness 
\end{itemize}

However, the method itself has some drawbacks. {\tt{CALL}} instructions can be encoded in various ways (relative, absolute, indirect), which can influence the scanner's performance and also the potential reliability of this method. Secondly, only original return instructions can be protected. In other words, using different instructions (like indirect jumps or calls) for linking gadgets will be not detected. On the other hand, the {\tt{CALL}} opcode checking method can be based on opcode-frequency statistics, which could decrease the potential performance slowdown. Additionally,  since only specified (valuable for attacker) return instructions can be protected, this should have a positive influence for the program's performance.

\subsubsection{Emitting magic values}

This method was introduced by the Pax Team~\cite{PaxFuture} and it relies on emitting magic bytes after every {\tt{CALL}} instruction and testing them at the function epilogue, as shown in Listing~\ref{pax_method}.

\begin{figure}

{\ttfamily{\footnotesize{
\lstset{language={[x86masm]Assembler}}
\begin{lstlisting}[frame=trbl, label=pax_method, caption={Protection of the execution flow changes via the return instructions.}, captionpos=b]{}
callee
epilogue:
	mov register,[esp]
	cmp [register+1],MAGIC
	jnz .1
	retn
.1: jmp esp

caller:
	call callee
	test eax,MAGIC
\end{lstlisting}
}}}
\end{figure}

This method seems to be more reliable than the method described in Section~\ref{testing_calls}, although it also has some major drawbacks. First of all, the {\tt{TEST}} instruction isn't neutral for the application context's state, since the {\tt EFLAGS} register is modified by this instruction. This flaw\footnote{Whether this is a flaw or not depends mostly on the application binary interface; in most cases the caller is responsible for saving the flags.}, however, can be easily fixed by simply emitting {\tt{JMP OVER\_MAGIC}} instruction after each {\tt{CALL}}. A more serious limitation of this method is the fact that every module used by the application would have to be created (compiled and linked) with the same {\tt{MAGIC}} value. This is necessary since execution transfers may occur from one module to another\footnote{Modules that don't perform execution transfers to other modules can be left ``unsynchronized''.}.

Since this approach would be almost impossible to implement in the real world there is another solution which can be used here. We propose that Windows' Portable Executable loader be responsible for synchronizing every {\tt{MAGIC}} value after each system boot (and after specified module is loaded). This would of course require creating a new section (or some new, specific data directory) with all the {\tt{MAGIC}} values offsets that should be updated by the executable loader.

\subsection{Obfuscating instructions}

This approach addresses the problem where the {\tt{RET}} instruction opcode is a part of different instruction (typically it is located among the first 1-3 bytes, not including instruction opcode). Owing to our tests and external sources~\cite{RopPaper} most of such opcodes are found in the ModR/M byte. A second large source of {\tt{RET}} opcodes is found in immediate displacements. In order to prevent from effectively using such cases in the ROP attack we propose that every instruction with {\tt{RET}} opcode inside of its body will be obfuscated in a special manner. Of course control transfer instructions or any other instructions that use immediate data offsets are an exception to this rule since the immediate displacements are calculated by the linker. The potential obfuscation can be done in following fashion:

\begin{itemize}
    \item If {\tt{RET}} opcode is found in the first byte after the original instruction opcode, a jump land should be emitted just before this instruction. Such jump land should consist of a short unconditional jump instruction and a land (up to 16 bytes) of {\tt{INT3}} or other worthless for attacker single byte instructions. Such emitted instructions will never be executed by the original program flow because of the unconditional jump, which transfers the execution directly to the potentially dangerous instruction. Such action should decrease the number of effective gadgets used for creating the ROP chain.
    \item If {\tt{RET}} opcode is spotted in immediate constant values such instruction should be obfuscated for example by splitting {\tt{ADD REG,IMM32}} into two {\tt{ADD}} instructions where the {\tt{IMM32}} operand for both of them would be free of return instruction opcodes. Of course special care must be taken regarding the {\tt{EFLAGS}} register state after each such transition.
    \item If {\tt{RET}} opcode is found in ModR/M byte, which indicates using {\tt{EAX}} register as destination operand and {\tt{EBX}} register as source operand (e.g., {\tt{MOV EAX,EBX}}), such instructions can be transformed into equivalent form which doesn't include return instruction opcodes. For example {\tt{MOV EAX,EBX}} $\leftrightarrow$ {\tt{PUSH EBX; POP EAX}} ({\tt{0x53 0x58}}).
\end{itemize}

As previously mentioned, the presented solutions can only be applied to instructions that do not use immediate displacements, as those are handled by the linker.

\section{Binary-level mitigations}
\label{sec:binary_level}

In this section we present mitigations against ROP attacks that can be applied without any information of the program's original source code. All mitigations included in this section can be implemented at the binary-level.

\subsection{Stack Encapsulation}

To make a ROP attack work, the attacker must be able to point the stack pointer into the controlled data. In typical stack-buffer overflow vulnerabilities this is not needed, but in other vulnerabilities (e.g., heap-overflow) this is often a must. In order to achieve this goal, the attackers use the so called stack pivot sequence~\cite{ROPDino}. Listing~\ref{pivot_sequences} shows some commonly used stack pivot sequences. Our mechanism tries to take advantage of this information.

\begin{figure}

{\ttfamily{\footnotesize{
\lstset{language={[x86masm]Assembler}}
\begin{lstlisting}[frame=trbl, label=pivot_sequences, caption={Typical stack pivot sequences.}, captionpos=b]{}
1: mov esp,eax
   ret

2: xchg eax,esp
   ret

3: add esp,<number>
   ret
\end{lstlisting}
}}}
\end{figure}

When a new thread is created, operating systems reserve some necessary space for its stack memory. Stack borders are described in the {\tt{INITIAL\_TEB}} structure which is passed in one of parameter of {\tt{NtCreateThread}} function. Additionally stack borders are also available in the Thread Information Block ({\tt{FS:[0x04]}} - top stack, {\tt{FS:[0x08]}} - current bottom stack). When the attacker uses the pivot sequence he typically exceeds the stack border limits set by the thread initialization procedure. The methods described in the following sections were designed to recognize this behavior. Similar support must be taken when dealing with fibers, since they also use separate stacks.

\subsubsection{New stack segment descriptor}
\label{ldt_stack}

Microsoft Windows systems allow usermode applications to create their own local descriptor table (LDT). Most current operating systems use the flat memory model, where there is no need to create additional segments for every running application. This would be in fact a step back to the old segmented memory model. On Windows platforms, in usermode, all segments' base addresses are equal to zero, except the one pointed by the {\tt{FS}} register (the {\tt{GS}} segment register is not used\footnote{This is true for x86-32 architectures only}). In our mitigation mechanism we have developed two approaches that protect the system against the stack pivoting technique. Our initial technique was to create a stack segment descriptor each time new a thread is created with a base address equal to the stack bottom and limit corresponding to stack size. After the new segment is created we initialize the {\tt{SS}} segment register with a new value.

This method however has a big drawback, which is explained on the listing below (Listing~\ref{stack_drawback}).

\begin{figure}

{\ttfamily{\footnotesize{
\lstset{language={[x86masm]Assembler}}
\begin{lstlisting}[frame=trbl, label=stack_drawback, caption={Typical program instructions.}, captionpos=b]{}
xor eax,eax
lea edi,[esp+VALUE]
stosd
stosd
...
\end{lstlisting}
}}}
\end{figure}

The {\tt{LEA}} instruction is responsible for initializing the {\tt{EDI}} register with the effective address of {\tt{ESP+VALUE}}. However, the value that will be stored in the {\tt{EDI}} register is still relative to the stack segment base address (which is not null in our case). The problems start with instructions that don't use the {\tt{SS}} segment register for addressing purposes. For example, the {\tt{STOSD}} instruction uses the {\tt{ES}} segment register; its execution will end with an access violation, since the base address of the segment pointed by {\tt{ES}} segment register is different. In other words the {\tt{LEA}} instruction does not honor the segment registers when calculating the effective address.

To resolve this issue we were forced to change the base address of the newly created stack segments. To avoid unnecessary access violations, the stack segment base address was set to zero and its limit was set to the stack's top value. This has some disadvantages, since the attacker would need to initialize new stack pointer value with an address higher than the segment limit to trigger the mechanism. Most of the time, however, newly allocated buffers have higher addresses since the thread's stack memory allocation was done earlier (there are a few obvious exceptions to this rule). Each time the attacker tries to exceed the boundaries of the current stack segment a general protection fault occurs and, at this point, our filtering procedure decides if the selected process is being exploited and needs to be terminated.

As a side note, there is one small problem with this method. Instructions that use the {\tt{EBP}} register for memory addressing are also using the stack segment specified by the current segment selector. This means that if the {\tt{EBP}} is not related to the stack memory and the destination address exceeds the stack segment boundaries a general protection fault will occur. Such cases however can be easily filtered and the execution can be resumed after emulating the faulting instruction.

\paragraph{Countermeasures}

In order to bypass the stack encapsulation protection, an attacker would need to initialize the stack pointer with a lower memory address than the stack's top value. For example attacker can heap-spray the memory and then cause the application to create a new thread that will be used to trigger the vulnerability. By doing this attacker fake stack will be below the stack base. Another way would be to execute a gadget that reinitializes the stack segment with the original value (constant between Windows versions) by, for example, executing a {\tt{POP SS}} instruction. To disable this attack we are constantly monitoring the value of the {\tt{SS}} segment register, and we reinitialize it every time execution returns from a system call (since kernel reinitializes the segment registers values before the control is returned to the usermode).

\subsubsection{Monitoring stack pointer changes}

Another approach for detecting the stack pivoting technique is to monitor the stack pointer value at crucial areas. For example, instead of setting another segment for stack space we can hook important offensive API functions (e.g., {\tt{VirtualAlloc}}, {\tt{VirtualProtect}}) and test the stack pointer value there. Obviously, there is no guarantee that the attacker wouldn't be able to restore the original stack pointer before using such API functions. To improve the security level of this protection mechanism we also propose that newly allocated memory regions (or memory regions with changed page protection rights) with executable pages should be marked as non-executable\footnote{this requires having a CPU with NX bit support}. Now the page marked as non-executable by our mechanism will work as a decoy. If the processor is trying to execute the non-executable page (page protection was previously changed by our mechanism) then we firstly apply our filtering procedure which tests the stack pointer value. If everything is correct, the executable rights are re-enabled and the execution is continued --- the entire mechanism works like a one-time decoy.

\subsection{Code Encapsulation}
\label{code_encap}

The ROP technique, just like any other, has some strategic points. One of those is the fact that the attacker must know the virtual addresses of the used gadgets. Because so, the ASLR mechanism successfully obstructs the exploitation process. However, in some cases the attacked application is either not compatible with ASLR or just uses external modules which do not support the ASLR mechanism. There are also cases where the attacker is able to leak or guess the wanted virtual address, rendering the ASLR mechanism relatively easy to bypass. In this section we present a mechanism which will take advantage of this (ROP) technique's weak spot.

\begin{figure}[tbhp]
\centering
\includegraphics[scale=0.4]{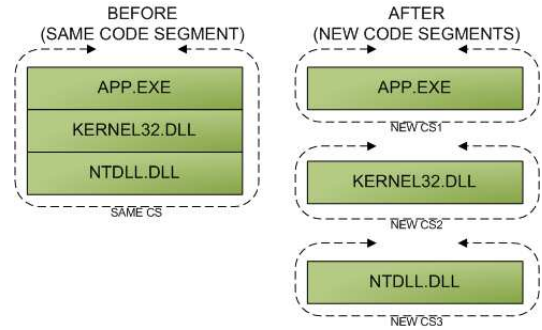}
\caption{Not-encapsulated and encapsulated modules inside of the process memory.}
\label{img:sep_seg}
\end{figure}

As stated in Section~\ref{ldt_stack}, Windows systems allow usermode applications to create their own local descriptor tables. In this mechanism we propose that each loaded module's code in the application's address space (including the main module) will have a separate segment for code sections\footnote{this mechanism is a bit similar to PaX SEGMEXEC~\cite{PaxSeg}}, as~\autoref{img:sep_seg} shows. Each time a execution transfer between modules or execution transfer using a full virtual address (including module imagebase value) occurs, a general protection fault will happen. At this point the filtering procedure decides whether this execution transfer attempt is valid or an attack attempt.

{\noindent\newline}This method has some drawbacks:

\begin{itemize}
    \item {A lot of control transfers are done through API calls and since they require a code segment switch a general protection fault is thrown every time such action occurs. Since this has a negative impact on the application's performance, entire import address table entries should be redirected to specific API stubs as shown in Listing~\ref{api_stub}.

        This solution should successfully decrease the negative performance impact because of the decreased number of GP faults. However, this is only one aspect of the problem, since the requested API must be able to return correctly to the specified location which is outside the current code segment. There are a few ways to solve this issue. One of the potential solutions can be based on faking the return address in the API land stub and then recalculating the correct address when {\tt{RET}} instructions cause the GP fault. Every potential solution here, however, will decrease the program performance. Additionally, since API lands can be only generated either before the base address of the specified module or attached to the end of it the code segment borders need to be expanded as well (at least in cases that don't overwrite module's memory).}

    \item {Special care must be taken when dealing with case-switch offsets since they also contain virtual addresses that don't apply to the new code segment limits. This issue can be partially resolved with using module relocation information and applying some heuristic scanning mechanism. All case-switch offsets found should be recalculated again and now point to relative addresses. However, some modules do not provide relocation information which makes dealing with such cases hard and probably slow.}

    \item {Some Portable Executable modules like {\tt{SHELL32.DLL}}  are pretty large (8MB - 16MB). This causes some additional problems, since if a function callback address located in different module has a virtual address somewhere between this 0-X MB range and some instruction will try to execute this virtual address the mechanism will fault. This is caused because the function's callback virtual address is located within the limits of the current code segment, and therefore the GP fault does not occur. This is a major drawback since it will likely lead to an application crash. Potential workarounds for this issue would be to disallow (or reserve) the memory located at the 0-X MB range. However, this would require an interaction with the system's Portable Executable loader.}

    \item {As explained before, every time kernel returns control to the usermode code segment registers are reinitialized with default values. Thus, the protection mechanism needs to re-initialize them as well each time such action happens.}

    \item {Additionally the number of segment descriptors is limited however this is not a problem for most of the applications since the number of loaded modules is not high.}

    \item {Special care must be taken when dealing with original code hooks, since such cases exists in some of the applications (for example in {\tt{IEXPLORE.EXE}} this is done by the {\tt{IEFRAME.DLL}} module).}

\end{itemize}

\begin{figure}

{\ttfamily{\footnotesize{
\lstset{language={[x86masm]Assembler}}
\begin{lstlisting}[frame=trbl, label=api_stub, caption={Example implementation of IAT redirection.}, captionpos=b]{}
CALL DWORD PTR DS:[0x406010]

original memory at 0x406010:
(ptr to user32.CreateWindowExA)
00406010 A9 E4 37 7E

patched memory at 0x406010:
00406010 dd offset api_land1

api_land1:
jmp user32_cs:rel_offset
\end{lstlisting}
}}}
\end{figure}

\paragraph{Countermeasures} The attacker would have to restore the original {\tt{CS}} register value by, for example, returning into a {\tt{RETF}} instruction. To protect against such attacks, firstly the current {\tt{CS}} segment will be monitored at the crucial program places and secondly all the newly generated code segment selector values will be pseudo-randomized.

\subsection{Code Decoys}

\begin{figure}[tbhp]
\centering
\includegraphics[scale=0.45]{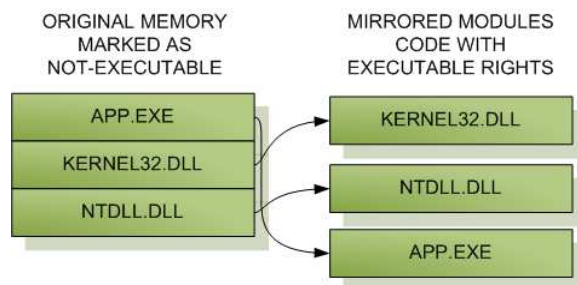}
\caption{Code decoys created from original modules.}
\label{img:code_decoys}
\end{figure}

This approach requires a processor with NX bit~\cite{NXArticle} support. The method itself is rather simple and it can be described in few steps:

\begin{enumerate}
    \item Mechanism setups a page fault filter and also module filter, which activates each time after a new module is mapped into process memory.
    \item All code sections from selected module found in the process memory are relocated to random memory address with the preservation of the section alignment (see \autoref{img:code_decoys}).
    \item After the relocation is done original code sections are marked as not-executable  (see \autoref{img:code_decoys}).
    \item Each time a page fault occurs because of an execution attempt of not-executable memory, the filtering procedure decides if it should recalculate the instruction pointer and continue the execution or to kill the process because of exploitation attempt.
\end{enumerate}

A similar idea was also used by the PaX Team in RANDEXEC mechanism and also by Matt Miller in the WehnTrust project~\cite{WehnTrust}.

To improve the performance of this mechanism, special care must be taken when dealing with case-switch offset tables --- this was already mentioned in Section~\ref{code_encap}. Additional performance improvements can be achieved with import address table redirecting, not unlike the idea explained in Section~\ref{code_encap}. It is important, however, to point out that this idea can also lower the protection level of this mechanism.

\paragraph{Countermeasures} The attacker would have to guess or leak the mirrored code address.

\section{Acknowledgments}

Author would like to thank Brad Spengler, Matt Miller and the Kryptos Logic team for helping with this article.

\section{Conclusion}

%The formal exploration of additional ideas behind the filtering procedure is left to the reader to deliberate. 

In this article, a number of promising techniques which can be used against the return-oriented programming attacks were presented. 

Most of the implementation problems of such mitigations are directly linked to a heavy performance impact. This is also a major factor in discouraging incorporation of these (and other) ROP mitigations into the selected platforms. Our security mitigations do not solve the problem of using return-oriented programming attacks completely, but they can effectively trammel and limit their usage. 

\newpage
\bibliographystyle{plain}
\bibliography{bibliografia}
\end{document}